\documentclass{elsart}

\usepackage{natbib}

\usepackage{epsfig}

\usepackage{amssymb}

\begin{document}

\begin{frontmatter}



\title{
Influence of Dark Matter on Light Propagation
in Solar System
}

\author{Hideyoshi ARAKIDA\corauthref{cor}}
\address{
School of Education, Waseda University\\
1-6-1, Nishi-Waseda, Shinjuku, Tokyo 169-8050, Japan
}
\corauth[cor]{Corresponding author}
\ead{arakida@edu.waseda.ac.jp}

\begin{abstract}
We investigated the influence of dark matter on light propagation in 
the solar system. We assumed the spherical symmetry of spacetime 
and derived the approximate solution of the Einstein equation, which 
consists of the gravitational attractions caused by the central 
celestial body, i.e. the Sun, and the dark matter surrounding it. 
We expressed the dark matter density in the solar system 
in the following simple power-law form, 
$\varrho(t, r) = \rho(t)(\ell/r)^k$, where $t$ is the coordinate time; 
$r$, the radius from the central body; $\ell$, the normalizing 
factor; $k$, the exponent characterizing $r$-dependence of 
dark matter density; and $\rho(t)$, the arbitrary function 
of time $t$. On the basis of the derived approximate solution, 
we focused on light propagation and obtained the additional corrections 
of the gravitational time delay and the relative frequency shift 
caused by the dark matter. As an application of our results, we 
considered the secular increase in the astronomical unit reported 
by Krasinsky and Brumberg (2004) and found that it was difficult to 
provide an explanation for the observed 
$d{\rm AU}/dt = 15 \pm 4 ~[{\rm m/century}]$.
\end{abstract}


\end{frontmatter}

\parindent=0.5 cm

\section{Introduction}
The existence of dark matter was first indicated by \citet{zwicky1933} 
and subsequently by \citet{rubin1970,rubin1980}. According to the recent 
cosmological observation, i.e. Wilkinson Microwave Anisotropy Probe 
(WMAP) \citep{wmap}, it is suggested that 
the majority of mass in our Universe is the dark matter, which is 
approximately 6 times that of ordinary (baryonic) matter. Furthermore, 
it is considered that dark matter has played an important role 
in the large-scale structure formation of the Universe. 
Actual observations such as 
2dF\footnote{\tt http://www.aao.gov.au/2df} and 
SDSS\footnote{\tt http://www.sdss.org/}
and results of numerical simulations, which assume the existence of 
dark matter such as Virgo 
Consortium\footnote{\tt http://www.virgo.dur.ac.uk/},
are in good agreement with each other. Therefore, dark matter is 
considered to be the fundamental material of our Universe, 
even though its details are presently cloaked in mystery.

If dark matter is an essential component in the Universe,
it is interesting and worthy to investigate its
existence and detectability in our neighborhood area, the solar system.
The number of recent astronomical and astrophysical measurements in 
the solar system have drastically increased, and they have led to 
(i) deep understanding of planetary dynamics and fundamental 
gravitational physics, (ii) significant improvement in lunar and 
planetary ephemerides, and (iii) precise determination of various 
astronomical constants.

To date, the gravitational influence of dark matter on planetary 
motion, such as the additional perihelion advance, has 
been studied by several authors. For instance, 
the effect of galactic dark matter has been considered by
\citet{braginsky1992}, \citet{klioner1993}, \citet{nordtvedt1994},
and \citet{nordtvedt1995}.
On the other hand, the upper limit of the dark matter density 
in the solar system has been estimated within a range
\citep{anderson1989,gron1996,khriplovich2006,iorio2006,sereno2006,khriplovich2007,frere2008}
such that
\begin{eqnarray}
\rho_{\rm dm}^{\rm (max)} < 10^{-16} \sim 10^{-20} ~{\rm [g/cm^3]}.
\label{estm-dm}
\end{eqnarray}
Recently, the upper limit of the planet-bound dark matter was 
also evaluated \citep{adler1,adler2}.

However, its contribution to light propagation 
has hardly been examined, in spite of the fact that the current 
accurate observations have been archived by an improvement in the
observation of light/signal; round-trip time 
(radar/laser gauging techniques and spacecraft ranging), 
amelioration of atomic clocks, radio links of spacecraft, 
and increasing stability of frequency standard. 
Moreover, planned space missions, such as 
GAIA\footnote{\tt http://www.rssd.esa.int/index.php?project=GAIA\&page=index}, 
SIM\footnote{\tt http://planetquest.jpl.nasa.gov/SIMLite/sim\_index.cfm}, 
LISA\footnote{\tt http://lisa.nasa.gov/}, 
LATOR \citep{lator}, and ASTROD/ASTROD-1 \citep{ni2007},
require accurate light propagation models. Therefore, it is 
noteworthy to examine how dark matter in the solar system affects 
light propagation and whether its traces can be detected
if it really exists in our solar system.

On the other hand, because of a drastic improvement in the 
measurement techniques used in the solar system, some 
unexplained problematic phenomena occurred, such as 
pioneer anomaly \citep{anderson2}, 
Earth flyby anomaly \citep{anderson3}, secular increase in the
astronomical unit \citep{kb2004}, and anomalous perihelion 
precession of Saturn \citep{iorio2009}. 
Currently, the origins of these phenomena are far from clear, 
nevertheless, dark matter may cause some significant contribution 
to these phenomena \citep{nieto2008,an2009,adler3}.

In this study, we will examine the influence of dark matter on light 
propagation in the solar system. First, we assume the spherical 
symmetry of spacetime and derive a simple approximate solution of 
the Einstein equation, which consists of gravitational attractions 
caused by the central celestial body, i.e, the Sun, and the dark 
matter surrounding it. Then, we will focus on formulating a light 
propagation model and estimating the additional effects of gravitational 
time delay and the relative frequency shift of a signal.
As an application of our results, we will consider the secular 
increase in the astronomical unit (of length), AU, 
reported by \citet{kb2004}.

This paper is organized as follows. In Section \ref{model}, we 
explain the model of spacetime and some assumptions. In Section 
\ref{solution}, we derive the approximate solution of the Einstein 
equation. In sections \ref{timedelay} and \ref{frequency}, we investigate 
time delay and relative frequency shift, respectively.
In section \ref{application}, we focus on the application of our 
results to the secular increase in the astronomical unit. 
Finally, in section \ref{summary}, we provide the summary of
our study.
\section{Model and Assumptions\label{model}}
Before deriving the approximate solution of the
Einstein equation, we explain the model of spacetime and some 
assumptions. First, we suppose that the spacetime is 
characterized by the gravitational attractions caused by the
central celestial body, i.e. the Sun, and dark matter surrounding it.
Then, we express the spherically symmetric form of metric as
\begin{eqnarray}
 ds^2 = -e^{\mu} c^2dt^2 + e^{\nu}dr^2 + 
  r^2(d\theta^2 + \sin^2\theta d\phi^2),
  \label{metric1}
\end{eqnarray}
where $e^{\mu}$ and $e^{\nu}$ are functions of time $t$ and 
radius $r$, and $c$ is the speed of light in vacuum.

As the stress-energy tensor $T^{\alpha}_{\beta}$
\footnote{In this paper, Greek indexes run from 0 to 3, and 
Latin ones do from 1 to 3.},
we presume the following form
\begin{eqnarray}
 T^0_0 = -\varrho(t, r)c^2,\quad
  T^1_0 = \sigma(t, r)c,\quad 
  T^i_j = 0.
  \label{s-e-tensor1}
\end{eqnarray}
$T^0_0$ is related to the dark matter density $\varrho(t, r)$.
At this time, since we do not have any knowledge about the 
actual distribution of dark matter in the solar system, 
then we suppose the following simple power-law form:
\begin{eqnarray}
 \varrho(t, r) = \rho(t)\left(\frac{\ell}{r}\right)^k,
\end{eqnarray}
where $\ell$ is the normalizing factor that is chosen as 
$\ell \equiv r_{\rm E}$ and $r_{\rm E}$ is the orbital radius of 
Earth, $k$ is the exponent characterizing $r$-dependence of 
$\varrho(t, r)$, and $\rho(t)$ represents the time variation of the 
dark matter density\footnote{$\rho(t)$ can be considered to be the
dark matter density observed around the Earth's orbit since 
$\ell = r_{\rm E}$.}. The test particle, i.e. photon, is subjected 
to the gravitational attractions caused by the central body and 
dark matter, which is confined within a spherical shell of radius $r$ at  
time $t$. $\rho(t)$ is an arbitrary function of time $t$, 
however, for the sake of simplicity, we assume that the 
time variation of the dark matter density in the solar system 
is considerably slow;
\begin{eqnarray}
\rho(t) \simeq \rho_0 + 
\left.\frac{d\rho}{dt}\right|_0 (t - t_0),
\label{rho-def}
\end{eqnarray}
where subscript 0 denotes the initial epoch of planetary ephemerides.
As the one possibility, the time variation of the dark matter density 
$\varrho(t, r)$ may be caused by the motion of the solar system 
in our Galaxy, if the distribution of galactic dark matter 
is inhomogeneous.

The dark matter density $\varrho(t, r)$ observed in the solar system 
may be generally expressed as a sum of solar system-bound 
(or local) dark matter $\varrho^{\rm (solar)}(t, r)$ and galactic 
dark matter $\varrho^{\rm (galactic)}(t, r)$ as follows:
\begin{equation}
 \varrho(t, r) = \varrho^{\rm (solar)}(t, r) + 
  \varrho^{\rm (galactic)}(t, r).
\label{dm-density}
\end{equation}
Here, to simplify the situation, we assume that the spacetime is
spherically symmetric and the time variation of the
dark matter density $\varrho(t, r)$ is caused by the inhomogeneity of 
the galactic dark matter as mentioned above. Therefore it is 
possible to express
\begin{eqnarray}
 \varrho^{\rm (solar)}(t, r) &=& \varrho^{\rm (solar)}(r) = 
  \rho_0^{\rm (solar)}\left(\frac{\ell}{r}\right)^k,\\
  \varrho^{\rm (galactic)}(t, r)
  &=& \left[
     \rho_0^{\rm (galactic)} +
     \left.\frac{d\rho^{\rm (galactic)}}{dt}\right|_0
     (t - t_0)
    \right]\left(\frac{\ell}{r}\right)^k.
\end{eqnarray}
In this case, $\rho_0$ and $d\rho/dt|_0$ in (\ref{rho-def}) are
\begin{equation}
 \rho_0 = \rho_0^{\rm (solar)} + \rho_0^{\rm (galactic)},\quad
  \left.\frac{d\rho}{dt}\right|_0 = 
  \left.\frac{d\rho^{\rm (galactic)}}{dt}\right|_0.
\end{equation}
According to the recent investigation, i.e. \citet{bm2005}, 
the galactic dark matter density is of the order of 
$10^{-24} ~[{\rm g/cm^3}]$. This is several orders of magnitude 
smaller than the evaluated density of dark matter in the solar 
system (see (\ref{estm-dm})). Therefore in this study, we suppose
that
\begin{equation}
\rho_0 = \rho_0^{\rm(solar)}.\quad
  \left.\frac{d\rho}{dt}\right|_0 = 
  \left.\frac{d\rho^{\rm (galactic)}}{dt}\right|_0.
\end{equation}
$T^1_0, T^0_1$ represent the time variation of energy and momentum 
flux. We auxiliary introduced these components to preserve the 
time dependency of the obtained solution. Though $T^i_j$ 
represents the stress part attributed to the dark matter, 
currently, the equation of state of dark matter is not known, 
therefore, we adopt the standard assumption that the dark matter 
is pressure-less dust particles $p \simeq 0$ and that its time 
variation is also negligible $dp/dt \simeq 0$.

\begin{figure}
   \begin{center}
    \includegraphics[scale=0.65]{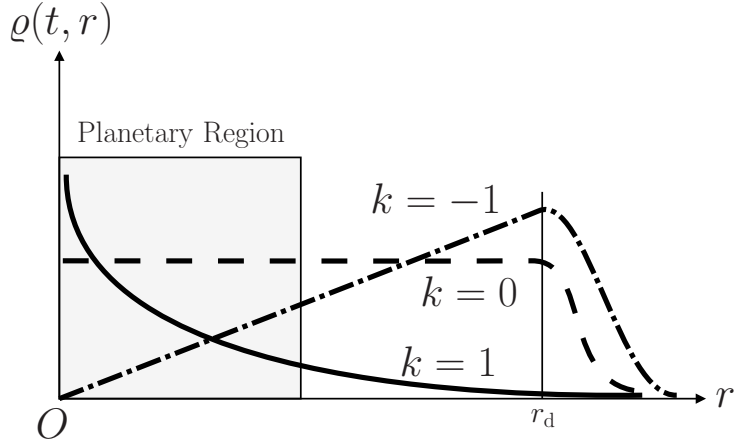}
   \end{center}
   \caption{Conceptual diagram of $r$-dependence of dark matter
   density for $k=1$, 
   $k = 0$, and $k = -1$.  
   We simply assume that the actual observations are carried out within 
 the quite inner (planetary) area of the solar system
   $\ll r_{\rm d}$.
    \label{fig1}}
  \end{figure}

Finally, we consider the choice of exponent $k$ of $\varrho(t, r)$. 
Because the distribution of dark matter in the solar system 
is poorly understood, we adopt the following three indexes as 
examples: 
$k = 1$ (density decreasing with $r$), $k = 0$ (constant density), 
and $k = -1$ (density increasing with $r$). The density $\varrho(t, r)$ 
should be damped at a certain radius $r_{\rm d}$ from the Sun 
(especially when $k = 0$ and $k = -1$), and it is natural 
to imagine that $\varrho(t, r)$ reaches asymptotically for the 
galactic dark matter density, i.e. $\sim 10^{-24}~ [\rm g/cm^3]$. 
However, we are now interested in the astronomical observations within 
the quite inner (planetary) area of the solar system (see Fig. \ref{fig1} 
for the conceptual diagram). Therefore, in this study, 
we do not consider the details of 
dark matter density distribution far away from the Sun. 
\section{Approximate Solution of the Einstein Equation\label{solution}}
On the basis of the assumptions in Section \ref{model}, 
we obtain the approximate solution of the Einstein equation
\begin{eqnarray}
G^{\alpha}_{\beta}\equiv
R^{\alpha}_{\beta} - \frac{1}{2} \delta^{\alpha}_{\beta}R = 
\frac{8\pi G}{c^4}T^{\alpha}_{\beta},
\end{eqnarray}
where $G$ is the Newtonian gravitational constant, $G^{\alpha}_{\beta}$ 
is the Einstein tensor, $R^{\alpha}_{\beta}$ is the Ricci tensor, 
$R$ is the Ricci scalar, and $T^{\alpha}_{\beta}$ is the stress-energy 
tensor in (\ref{s-e-tensor1}). The non-zero components of the 
Einstein tensor are expressed as follows:
\begin{eqnarray}
 G^0_0 &=& e^{-\nu}
  \left(\frac{1}{r^2} - \frac{1}{4}
   \frac{\partial \nu}{\partial r}\right) - \frac{1}{r^2},
  \\
 G^1_1 &=& e^{-\nu}
  \left(\frac{1}{r^2} + \frac{1}{r}
  \frac{\partial \mu}{\partial r}\right) - \frac{1}{r^2},
  \\
 G^2_2 &=& G^3_3 =
  \frac{e^{-\nu}}{2}
  \left[n
   \frac{\partial^2\mu}{\partial r^2} + 
   \frac{1}{2}\left(\frac{\partial \mu}{\partial r}\right)^2
   -\frac{1}{2}\frac{\partial \nu}{\partial r}
   \frac{\partial \mu}{\partial r}
   +\frac{1}{r}
   \left(
    \frac{\partial \mu}{\partial r} - 
    \frac{\partial \nu}{\partial r}\right)
  \right]
  \nonumber\\
 & &\qquad
  - \frac{e^{-\mu}}{2c^2}
  \left[
   \frac{\partial^2 \nu}{\partial t^2} + \frac{1}{2}
   \left(
   \frac{\partial \nu}{\partial t}
   \right)^2
   - \frac{1}{2}\frac{\partial \mu}{\partial t}
   \frac{\partial \nu}{\partial t}
  \right],
  \\
 G^0_1 &=&
  - \frac{1}{cr}e^{-\nu}\frac{\partial \nu}{\partial t},\quad
 G^1_0 =
  \frac{1}{cr}e^{-\mu}\frac{\partial \nu}{\partial t}.
\end{eqnarray}
From the $00$ component of the Einstein equation, we have
\begin{eqnarray}
 r(1 - e^{-\nu}) = m(t)
  + \frac{8\pi G}{c^2}\frac{\rho(t)\ell^k}{3-k}r^{3-k},
\label{00-1}
\end{eqnarray}
where $m(t)$ is an arbitrary function of time $t$, however, 
we choose $m(t)$ such that it reduces to the Schwarzschild 
radius $m(t) \rightarrow m = 2GM/c^2 = \mbox{constant}$ when 
$\varrho(t, r) = 0$. Therefore, we obtain
\begin{eqnarray}
 e^{-\nu}
  = 1 - \frac{2GM}{c^2 r} -
  \frac{8\pi G}{c^2}\frac{\rho(t)\ell^k}{3-k}r^{2-k}.
\label{00-2}
\end{eqnarray}
Using the 00 and 11 components of the Einstein equation, 
it follows
\begin{eqnarray}
 \frac{\partial \mu}{\partial r} + 
  \frac{\partial \nu}{\partial r}
  =
 \frac{8\pi G}{c^2}\frac{\rho(t)\ell^k}{r^{k-1}},
\label{00-11}
\end{eqnarray}
where we kept the ${\cal O}(c^{-2})$ order terms only on 
the right-hand side. We obtain the following equation by
integrating (\ref{00-11}) with respect to $r$, combing it with 
(\ref{00-2}), and omitting the ${\cal O}(c^{-4})$ and 
higher order terms:
\begin{eqnarray}
 e^{\mu} = 
  f(t)
  \left[
   1- \frac{2GM}{c^2 r} + \frac{8\pi G}{c^2}
   \frac{\rho(t)\ell^k}{(2 - k)(3-k)}r^{2-k}
  \right].
\label{00-3}
\end{eqnarray}
Although $f(t)$ is also an arbitrary function of time $t$,
we replace the time coordinate with $\sqrt{f(t)}dt \rightarrow dt$ and 
delete $f(t)$. Finally, we obtain
\begin{eqnarray}
 ds^2 &=& -
  \left(1 - \frac{2GM}{c^2r} + \frac{8\pi G}{c^2}
  \frac{\rho(t)\ell^k}{(2-k)(3-k)}r^{2-k}\right)c^2dt^2
  \nonumber\\
 & &\quad
  + 
  \left(
   1 - \frac{2GM}{c^2r} - \frac{8\pi G}{c^2}
   \frac{\rho(t)\ell^k}{3-k}r^{2-k}
  \right)^{-1}dr^2
 + r^2 d\Omega^2,
\label{metric2}
\end{eqnarray}
where $d\Omega^2 = d\theta^2 + \sin^2 \theta d\phi^2$.
When $k = 0$ and $\rho(t) = \rho_0 = {\rm constant}$, 
(\ref{metric2}) reduces to the metric derived by \citet{gron1996}; 
therefore, our solution (\ref{metric2}) is considered to be some 
extension of the solution obtained by
Gr{\o}n and Soleng. In the case of a solar system experiment, 
it is sufficient to use the following approximate form:
\begin{eqnarray}
ds^2 &=& - \left[1 - \frac{2}{c^2}U(t, r)\right]c^2dt^2 + 
\left[1 + \frac{2}{c^2}V(t, r)\right]dr^2
+
r^2 d\Omega^2,
\label{metric3}
\end{eqnarray}
where
\begin{eqnarray}
U(t, r) &=&
\frac{GM}{r} - 
  \frac{4\pi G\rho(t)\ell^k}{(2-k)(3-k)}r^{2-k},\\
V(t, r) &=& 
\frac{GM}{r} +
   \frac{4\pi G\rho(t)\ell^k}{3-k}r^{2-k}.
\end{eqnarray}
We mention here that it is easy to incorporate the cosmological constant 
$\Lambda$ in (\ref{metric2}) or (\ref{metric3}). However, to focus on 
the effect of dark matter, we omit the $- \Lambda r^2/3$ term.
\section{Gravitational Time Delay\label{timedelay}}
\subsection{Time Delay in Coordinate Time}
In this section, we calculate the time delay attributed to the dark matter. 
In the static spacetime, we can easily relate an affine parameter 
$\lambda$ to coordinate time $t$ using by the Euler-Lagrange equation of 
$g_{00}$, i.e. Chapter 8 of \citet{weinberg}. 
However, (\ref{metric2}) or (\ref{metric3}) is the non-static or 
time-dependent then it is not easy to calculate the geodesic 
equation analytically in general. 
Therefore, we consider an alternative approach.

\begin{figure}
 \begin{center}
   \includegraphics[scale=0.65]{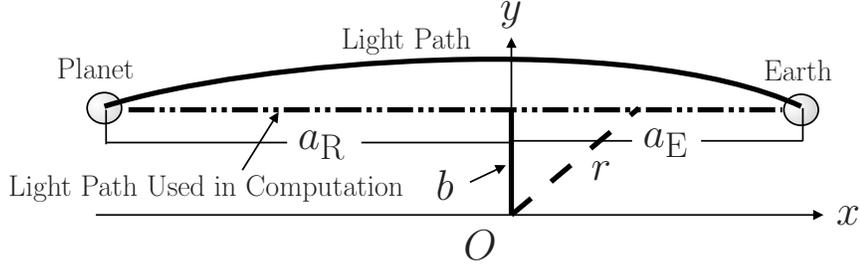}
 \end{center}
 \caption{Light/signal path. We assume that the
 first approximation of the light path is rectilinear 
 along the $x$-direction (bold dashed line),
 $b$ is the impact parameter, and $r = \sqrt{x^2 + b^2}$. 
 The actual light path is drawn by the bold solid line. 
 \label{fig2}}
\end{figure}

To begin with, we transform (\ref{metric3}) from spherical coordinates 
to rectangular coordinates. By the usual coordinate transformation, 
i.e.
\begin{equation}
x = r\sin \theta \sin \phi,\quad
y = r\sin \theta \cos \phi,\quad
z = r\cos \theta,
\end{equation}
(\ref{metric3}) is rewritten as (e.g., \citet{brumberg1991})
\begin{eqnarray}
ds^2 = -\left(1 - \frac{2}{c^2}U\right)c^2dt^2
+ \left(
\delta_{ij} + \frac{2}{c^2}V\frac{x^i x^j}{r^2}
\right)dx^i dx^j,
\label{metric4}
\end{eqnarray}
where $\delta_{ij}$ is the Kronecker's delta symbol. We suppose 
that the actual light path is calculated along with the approximate 
rectilinear path ($x$-direction) such that
\begin{equation} 
y = b = {\rm constant},\quad 
z = 0,\quad 
r = \sqrt{x^2 + b^2},
\end{equation} 
where $b$ is an impact factor (see Fig. \ref{fig2}).
Hence, (\ref{metric4}) becomes
\begin{eqnarray}
ds^2 = -\left(1 - \frac{2}{c^2}U\right)c^2 dt^2
+ \left(1 + \frac{2}{c^2}V\frac{x^2}{r^2}\right)dx^2.
\label{metric5}
\end{eqnarray}
The world line of the light ray is null geodesic $ds^2 = 0$, 
therefore from (\ref{metric5}) we have 
\begin{eqnarray}
c \frac{dt}{dx} 
=
1 + \frac{1}{c^2}
\left[
\frac{GM}{r}\left(1 + \frac{x^2}{r^2}\right)
- \frac{4\pi G \ell^k\rho(t)}{(2 - k)(3-k)}r^{2 - k}
+ 
\frac{4\pi G \ell^k\rho(t)}{3 -k}
\frac{x^2}{r^k}
\right].
\label{metric6}
\end{eqnarray}
We express $\rho(t)$ in the form of (\ref{rho-def}).

To obtain the round-trip time from (\ref{metric6}), we assume that
the lapse time $\Delta t$ is expressed by a linear combination of 
each effect as follows:
\begin{eqnarray}
\Delta t = \Delta t_{\rm pN} + 
\Delta t_{\rm dm}^{\rm (const)} + \Delta t_{\rm dm}^{(t)}
\end{eqnarray}
where $\Delta t_{\rm pN}$ corresponds to the Shapiro time delay 
in 1st post-Newtonian approximation, 
$\Delta t_{\rm dm}^{\rm (const)}$ is attributed
to the static part of the dark matter density ($\rho_0$ of 
(\ref{rho-def})), and $\Delta t_{\rm dm}^{(t)}$ is the contribution 
of the time-dependent part of dark matter density 
($d\rho/dt|_0t$ of (\ref{rho-def})).   
The post-Newtonian parts are easily obtained as follows:
\begin{eqnarray}
 \Delta t_{\rm pN} &=&\frac{x_2 - x_1}{c}
  \nonumber \\
 & &+
  \frac{GM}{c^3}
  \left[
   2 \ln \frac{x_2 + \sqrt{x^2_2 + b^2}}
   {x_1 + \sqrt{x^2_1 + b^2}}
   - 
   \left(
    \frac{x_2}{\sqrt{x^2_2 + b^2}} 
    - \frac{x_1}{\sqrt{x^2_1 + b^2}}
   \right)
  \right].
  \label{td-pn}
\end{eqnarray}
Next, we calculate the time delay caused by dark matter. The static part 
$\Delta t_{\rm dm}^{\rm (const)}$ is straightforwardly integrated as
\begin{eqnarray}
\Delta t_{\rm dm}^{\rm (const)} &=& \frac{\pi 
G \rho_0}{c^3} H(x_1, x_2; k)\\ H(x_1, x_2; k)
&=&
\left\{
\begin{array}{ll}
- 2\ell b^2 \ln \frac{x_2 + 
\sqrt{x^2_2 + b^2}}
{x_1 + \sqrt{x^2_1 + b^2}} & (k=1)\\
\frac{2}{3}
\left[
\frac{1}{3}(x^3_2 - x^3_1) - b^2(x_2 - x_1) 
\right] & (k=0)\\
- \frac{1}{12\ell}
\left[
3 b^4 
\ln \frac{x_2 + \sqrt{x^2_2 + b^2}}
{x_1 + \sqrt{x^2_1 + b^2}} \right.\\
\quad - 2 (x_2\sqrt{x^2_2 + b^2}^3 
- x_1\sqrt{x^2_1 + b^2}^3) \\
\quad + \left.
3b^2 (x_2\sqrt{x^2_2 + b^2} 
- x_1\sqrt{x^2_1 + b^2})
\right] & (k=-1).\\
\end{array}
\right.
\end{eqnarray}
Finally, we compute the time-dependent part
$\Delta t_{\rm dm}^{(t)}$ ($d\rho/dt|_0t$ part). If light is emitted 
from Earth at $t = T$ and it reaches the reflector (planet/spacecraft) 
at $t = T + \Delta t_{\rm dm}^{(t)}$, then $\Delta t_{\rm dm}^{(t)}$ 
satisfies
\begin{eqnarray}
c \int^{T + \Delta t_{\rm dm}^{(t)}}_T 
\frac{1}{t}dt
=
c \ln \frac{T + \Delta t_{\rm dm}^{(t)}}{T}
= \left.\frac{d\rho}{dt}\right|_0
\int^{x_2}_{x_1}\left[\cdots\right]dx,
\end{eqnarray}
where the integral $\int^{x_2}_{x_1}\left[\cdots\right]dx$ is the 
same as that in the case of the static part 
$\Delta t_{\rm dm}^{\rm (const)}$.
Since $d\rho/dt|_0 \ll 1$, $\Delta t_{\rm dm}^{(t)}$ can be
expressed as
\begin{eqnarray}
\Delta t_{\rm dm}^{(t)} = \frac{1}{c}
\left. \frac{d\rho}{dt}\right|_0 T
\int^{x_2}_{x_1}\left[\cdots\right]dx.
\end{eqnarray}
Therefore, the time delay caused by dark matter is expressed as
\begin{eqnarray}
\Delta t_{\rm dm} \equiv
\Delta t_{\rm dm}^{\rm (const)} + \Delta t_{\rm dm}^{(t)}
= \frac{\pi G }{c^3} 
\left(\rho_0 + \left.\frac{d\rho}{dt}\right|_0 T \right)
H(x_1, x_2; k).
\label{dt-dm}
\end{eqnarray}
In this case, the time delay caused by dark matter can be 
characterized by the density at the emission time of signal $t = T$, 
that is $\rho(T) = \rho_0 + d\rho/dt|_0T$.

In order to apply (\ref{td-pn}) and (\ref{dt-dm}) to 
two-way light propagation, i.e. 
Earth $\rightarrow$ receiver (planet/spacecraft) $\rightarrow$ Earth,
we consider the following situation; the light path used in computation 
is parallel to the $x$-axis, Earth, and the receiver (planet/spacecraft) 
are located $x = a_{\rm E}$ and $x = - a_{\rm R}$, respectively 
(see Fig. \ref{fig2} again). We suppose that (a) during the round-trip of 
light, Earth and the receiver are almost at rest and that (b) the time 
variation of the dark matter density is also considerably slow. 
In other words, 
the time lapse $\Delta T$ can be mainly determined using by the 
dark matter density at the emission time $t = T$, $\rho(T)$. Hence, 
the round-trip time in the coordinate time, $\Delta T$ is
expressed as 
\begin{eqnarray}
\Delta T 
&=&
2\frac{a_{\rm E} + a_{\rm R}}{c}+
\frac{2GM}{c^3}
\left[
2\ln \frac{(a_{\rm E} + \sqrt{a_{\rm E}^2 + b^2})
(a_{\rm R} + \sqrt{a^2_{\rm R} + b^2})}{b^2}
\right.\nonumber\\
& &-\left.
\left(
\frac{a_{\rm E}}{\sqrt{a^2_{\rm E} + b^2}} 
+ \frac{a_{\rm R}}{\sqrt{a^2_{\rm R} + b^2}}
\right)
\right]
\nonumber\\
& &+ 
\frac{2\pi G}{c^3}
\left(\rho_0 + \left.\frac{d\rho}{dt}\right|_0 T \right)
{\cal H}(a_{\rm E}, a_{\rm R}; k),
\label{time-delay1}
\end{eqnarray}
where we substitute ${\cal H}(a_{\rm E}, a_{\rm R}; k) 
= H(0, a_{\rm E}; k) + H(0, a_{\rm R}; k)$. To calculate
(\ref{time-delay1}), we referred to an approach in Section 40.4 and 
Figure 40.3 shown by \citet{mtw}.

\begin{figure}
   \begin{center}
   \includegraphics[scale=0.65]{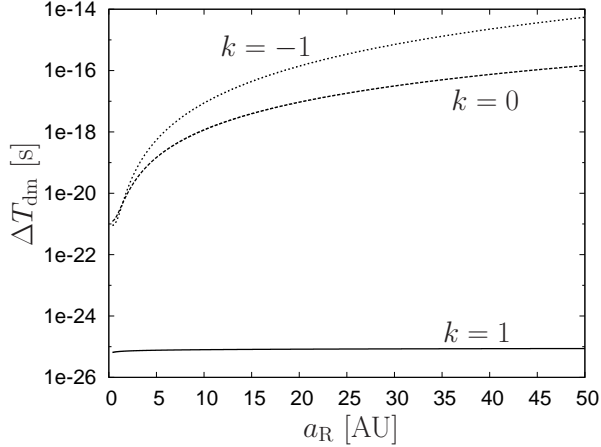}
   \end{center}
\vspace*{8pt}
   \caption{$a_{\rm R}$ dependence of additional time delay 
   $\Delta T_{\rm dm}$. As $\rho_0$, we adopt 
    $\rho_0 \sim 10^{-16} ~[{\rm g/cm^3}]$ and fixed
    $a_{\rm E} = 1.0 ~[{\rm AU}] (= 1.5 \times 10^{11} ~[{\rm m}])$ and
    $b = 0.001 ~[{\rm AU}] (= 1.5 \times 10^{8} ~[{\rm m}])$.
    \label{fig3}}
  \end{figure}

Let us estimate the order of the time delay $\Delta T_{\rm dm}$.
Because $d\rho/dt|_0 T$ is now anticipated to be
considerably smaller than the dominant part $\rho_0$, we neglect the
$d\rho/dt|_0 T$ term here and evaluate
\begin{eqnarray}
\Delta T_{\rm dm} \simeq
\frac{2\pi G \rho_0}{c^3}
{\cal H}(a_{\rm E}, a_{\rm R}; k).
\end{eqnarray}
Fig. \ref{fig3} illustrates the $a_{\rm R}$ dependence of the
time delay $\Delta T_{\rm dm}$. We adopted 
$\rho_0 \sim 10^{-16} ~[{\rm g/cm^3}]$, which is the largest upper limit 
obtained from the dynamical perturbation on planetary motion. We fixed 
$a_{\rm E} = 1.0 ~[{\rm AU}] (= 1.5 \times 10^{11} ~[{\rm m}])$ 
(orbital radius of the Earth) and impact parameter
$b = 0.001 ~[{\rm AU}] (= 1.5 \times 10^{8} ~[{\rm m}])$.
If the dark matter is accumulated in the neighborhood of the Sun 
($k = 1$), $\Delta T_{\rm dm} \sim 10^{-25} ~[{\rm s}]$ in a given 
range of $a_{\rm R}$. When $k = 0$ and $k = -1$, 
$\Delta T_{\rm dm}$ is of the order of $10^{-20} ~[{\rm s}]$ 
in the inner planetary region, while in the outer planetary
region, it is of the order of 
$10^{-19} < \Delta T_{\rm dm} < 10^{-17} ~[{\rm s}]$ 
($k = 0$) and $10^{-16} < \Delta T_{\rm dm} < 10^{-14} ~[{\rm s}]$
($k=-1$). However, the current observational limit in  
the solar system is $\sim 10^{-8} [{\rm s}]$ or a few 100 [m] for
planetary radar and $10^{-11} [{\rm s}]$ or a few [m] for 
spacecraft ranging; the internal error of the atomic 
clocks on Earth is $\sim 10^{-9} ~[{\rm s}]$. 
Then, at this time, it is difficult to extract the trace of dark matter 
from the ranging data.
\subsection{Time Delay in Proper Time}
The round-trip time of the light ray (\ref{time-delay1}) is 
expressed in the coordinate time. However, the actual measurement is 
performed by the atomic clocks on the surface of Earth, which shows 
proper time $\tau$. Therefore, we must transform (\ref{time-delay1}) into 
proper time. Presently, it is sufficient to use the equation of 
proper time for the quasi-Newtonian approximation such that
\begin{eqnarray}
\frac{d\tau}{dt} = 1 - \frac{1}{c^2}
\left(U + \frac{1}{2}v^2\right).
\end{eqnarray}
Evaluating $d\tau/dt$ around the orbit of Earth and keeping the 
${\cal O}(c^{-3})$ terms only, the round-trip time 
$\Delta \tau$ measured in proper time is given by
\begin{eqnarray}
\Delta \tau &=& \left.\frac{d\tau}{dt}\right|_E \Delta T\nonumber\\
&=&
2\frac{a_{\rm E} + a_{\rm R}}{c} +
\frac{2GM}{c^3}
\left[
2\ln \frac{(a_{\rm E} + \sqrt{a_{\rm E}^2 + b^2})
(a_{\rm R} + \sqrt{a^2_{\rm R} + b^2})}{b^2}
\right.\nonumber\\
& &-\left.
\left(
\frac{a_{\rm E}}{\sqrt{a^2_{\rm E} + b^2}} 
+ \frac{a_{\rm R}}{\sqrt{a^2_{\rm R} + b^2}}
\right)
\right]
+ 
\frac{2\pi G}{c^3}
\left(\rho_0 + \left.\frac{d\rho}{dt}\right|_0 T \right)
{\cal H}(a_{\rm E}, a_{\rm R}; k)\nonumber\\
& &
- 2\frac{a_{\rm E} + a_{\rm R}}{c^3}
\left[\frac{1}{2}v^2_{\rm E} + \frac{GM}{a_{\rm E}} - 
\frac{4\pi G \ell^k}{(2-k)(3-k)}a^{2-k}_{\rm E}
\left(\rho_0 + \left.\frac{d\phi}{dt}\right|_0 T\right)
\right],
\end{eqnarray}
where $v_{\rm E}$ is the orbital velocity of Earth.
\section{Relative Frequency Shift\label{frequency}}
\begin{figure}
   \begin{center}
   \includegraphics[scale=0.65]{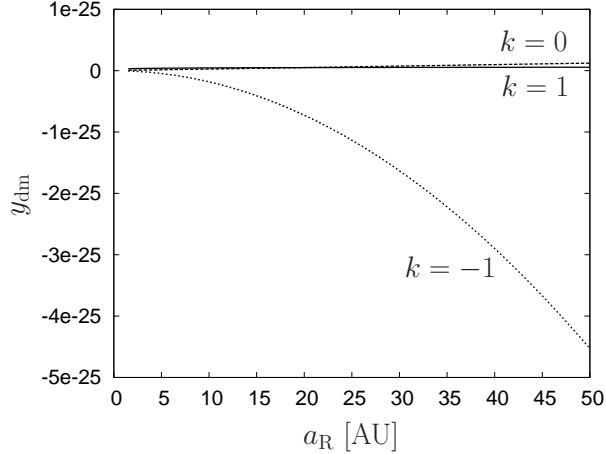}
   \end{center}
   \caption{Additional relative frequency shift caused by dark matter,
   and $y_{\rm dm}$ is plotted as a function of $a_{\rm R}$.
    We set $b_0 = 2 R_{\rm Sun}, R_{\rm Sun} 
    \simeq 6.9 \times 10^8 ~[{\rm m}], v \simeq 30 ~[{\rm km/s}]$,
    and $t = 1 ~[{\rm day}]$. 
    \label{fig4}}
  \end{figure}

We use (\ref{time-delay1}) to derive the relative frequency shift of 
signal $y$\footnote{Here, $y$ is not the $y$-coordinate, but the relative
frequency shift according to \citet{bertotti2003}.},
which is defined as
\begin{eqnarray}
 y = \frac{\delta \nu}{\nu} \equiv - \frac{d\Delta T}{dt}.
\end{eqnarray}
When the light ray passes near the limb of the Sun such as in the  
Cassini experiment \citep{bertotti2003}, the
conditions $a_{\rm E}, a_{\rm R} \gg b, da_{\rm E}/dt, {\rm and}
da_{\rm R}/dt \ll db/dt$ hold, where $b = \sqrt{b^2_0 + (vt)^2}$.
Then, the relative frequency shift caused by the Sun, $y_{\rm pN}$, 
and dark matter, $y_{\rm dm}$, are expressed as
\begin{eqnarray}
y &=& y_{\rm pN} + y_{\rm dm}\\
y_{\rm pN} &=& 
\frac{8GM}{c^3b}\frac{db}{dt}\\
y_{\rm dm} &=& \frac{\pi G}{c^3}
\left(\rho_0 + \frac{d\rho}{dt}T\right)
{\cal K}(a_{\rm E}, a_{\rm R}; k)\\
{\cal K}(a_{\rm E}, a_{\rm R}; k) &=&
\left\{
\begin{array}{ll}
8b\ell\left(
\ln \frac{4a_{\rm E}a_{\rm R}}{b^2} - 1 
\right)\frac{db}{dt} & (k=1)\\
\frac{8}{3}b(a_{\rm E} + a_{\rm R})\frac{db}{dt} & (k=0) \\
\frac{2}{\ell}
b\left[
b^2\ln 
\frac{4a_{\rm E}a_{\rm R}}{b^2}
- 
(a_{\rm E}^2 + a_{\rm R}^2)
\right]\frac{db}{dt} & (k=-1).\\
\end{array}
\right.
\end{eqnarray}
Fig. \ref{fig4} shows the relative frequency shift caused by
dark matter, $y_{\rm dm}$ as a function of $a_{\rm R}$. 
In this plot, we substitute $b_0 = 2 R_{\rm Sun}, R_{\rm Sun} 
\simeq 6.9 \times 10^8 ~[{\rm m}], v \simeq 30 ~[{\rm km/s}]$,
and $t = 1 ~[{\rm day}]$ ($t = 0$ gives the closest point). 
The order of magnitude of $y_{\rm dm}$ is 
$\sim 10^{-25}$; however, currently the stability of frequency 
standard is of the order of $10^{-15}$ or even higher. 
Therefore, the expected frequency shift caused by dark matter 
is approximately 10 orders of magnitude smaller than the present 
observational limit of frequency.
\section{Application to Secular Increase in Astronomical
Unit\label{application}}
In this section, we apply the previous results to the secular 
increase in the astronomical unit (of length) reported by \citet{kb2004}. 
The astronomical unit (AU) is one of the important scales in astronomy, 
and it is the basis of the cosmological distance ladder. 
AU is also a fundamental 
astronomical constant, which gives the relation between two length units; 
1 [AU] in the astronomical system of units and 1 [m] in SI ones. 
Presently, AU is determined by using the planetary radar and 
spacecraft ranging data (round-trip time of light ray), 
and the latest best-fit value is obtained as \citep{pitjeva2005}
\begin{eqnarray}
1 ~[{\rm AU}] = 1.495978706960 \times 10^{11} \pm 0.1 ~[{\rm m}].
\label{au-pitjeva}
\end{eqnarray}
We use the calculated planetary ephemerides (solution of equation of 
motion) to compute the theoretical value of the round-trip time 
$t_{\rm theo}$ using the following formula:
\begin{eqnarray}
 t_{\rm theo} = \frac{d_{\rm theo}}{c}{\rm AU} ~[{\rm s}],
\end{eqnarray}
where $d_{\rm theo} ~[{\rm AU}]$ is the interplanetary distance.
$t_{\rm theo} ~[{\rm s}]$ is compared with the observed
round-trip time $t_{\rm obs} ~[{\rm s}]$, and ${\rm AU}$ is 
optimized by the least square method. 

However, when Krasinsky and Brumberg replaced $t_{\rm theo}$ with
\begin{eqnarray}
t_{\rm theo} = \frac{d_{\rm theo}}{c}
\left[{\rm AU} + \frac{d{\rm AU}}{dt}(t - t_0)\right]
\end{eqnarray}
and fitted it to the observational data, they found that 
$d{\rm AU}/dt$ had a non-zero and positive secular value, 
$15 \pm 4 [{\rm m/century}]$, where $t_0$ is the initial epoch.
The evaluated value $d{\rm AU}/dt = 15 \pm 4 ~[{\rm m/century}]$
is approximately 100 times that of the current determination
error of ${\rm AU}$ (see (\ref{au-pitjeva})). 
At present, the time dependent part $(d\mbox{AU}/dt) (t - t_0)$
cannot be related to any theoretical predictions, 
hence, several attempts have been made to explain this secular 
increase in AU on the basis of various factors such as
the effects of cosmological expansion
\citep{kb2004,mashhoon2007,arakida2009}, mass loss of the Sun
\citep{kb2004,noerdlinger2008}, and time variation of the 
gravitational constant $G$ \citep{kb2004}.
However, unfortunately, thus far, none of these factors
seem to be responsible for the secular increase in AU.

It is noteworthy that the observed $d{\rm AU}/dt$ does
not imply the expansion of planetary orbit and/or an 
increase in the orbital period of a planet. As a matter of fact, 
the determination error of the latest planetary ephemerides is 
considerably smaller than the reported $d{\rm AU}/dt$ 
(see in Table 4. of \citet{pitjeva2005}). 
Hence, $d{\rm AU}/dt$ may be caused by some effects on light 
propagation, and not by the dynamical perturbation on 
planetary motion.

Moreover, AU denotes not only the conversion constant 
of the length unit but also the value that characterizes the
$GM$ of the Sun in SI units such that
\begin{equation}
GM_{\rm Sun} = k^2 {\rm AU}^3/{\rm d}^2~ [{\rm m^3/s^2}],
\end{equation}
where $k = 0.01720209895$ is the Gaussian gravitational constant,
and ${\rm d}$ is a day such that ${\rm d} = 86400~ [{\rm s}]$. 
Therefore, the observed $d{\rm AU}/dt$ may be related to
an increase in the dark matter density such that
$GM(t) = G(M_{\rm Sun} + M_{\rm dm}(t))$, where $M_{\rm dm}(t)$ is 
the total mass of dark matter within a planetary orbit at time $t$.

Then, let us evaluate the extent of time variation of
dark matter density $d\rho/dt|_0$ in (\ref{time-delay1}) 
that is needed to explain the observed $d{\rm AU}/dt$. We have
\begin{eqnarray}
 \frac{d_{\rm theo}}{c}\frac{d{\rm AU}}{dt}T 
 \sim
 \frac{2 \pi G}{c^3}
 \left.\frac{d\rho}{dt}\right|_0 T R^3,
\end{eqnarray} 
where we set ${\cal H}(a_{\rm E}, a_{\rm R}; k) \sim R^3$, and 
$R$ is the orbital radius of a planet. In the case of Earth--Mars ranging, 
we let $R \sim 1.52 ~[{\rm AU}]$ (orbital radius of Mars). 
To obtain the reported $d{\rm AU}/dt$, $d\rho/dt|_0$ must be  
of the order of $10^{-9} ~[{\rm g/(cm^3 s)}]$ and  
$d\rho/dt|_0 T \sim 1 [{\rm g/cm^3}]$ for $T \sim 100 ~[{\rm y}]$. 
However, this value corresponds to the density of water; therefore,
this possibility of achieving such value
is unrealistic and should be made an exception.
\section{Summary\label{summary}}
We investigated the influence of dark matter on light 
propagation in the solar system. We used the simplified model
to derive the approximate solution of the Einstein equation, which 
consists of the gravitational attractions caused by the central 
celestial body, i.e. the Sun, and dark matter surrounding it. 
We found that the derived metric (\ref{metric3}) can be considered 
to be an extension of the previous work by \citet{gron1996}. 
We assumed that the simple time variation of dark matter density, 
and focused our discussion on light propagation then computed 
the additional corrections of gravitational time delay and relative 
frequency shift. However, the expected effects were
considerably smaller than the
current observational limits, even when we considered the largest upper 
limit evaluated from the planetary perturbation 
caused by dark matter,
$\rho_0 \sim 10^{-16} ~[{\rm g/cm^3}]$.

We applied the obtained results to the secular increase in the 
astronomical unit reported by \citet{kb2004} and 
considered the possibility of explaining the observed 
$d{\rm AU}/dt = 15 \pm 4 ~[{\rm m/century}]$
on the basis of the time variation of the dark matter density.
We found that to induce the obtained $d{\rm AU}/dt$, the change in 
the dark matter density $d\rho/dt|_0$ in (\ref{rho-def}) must be 
of the order of $10^{-9} ~[{\rm g/(cm^3 s)}]$ and that 
$d\rho/dt|_0 T \sim 1 ~[{\rm g/cm^3}]$ for the interval
$T \sim 100 ~[{\rm y}]$. 
However, it is completely unrealistic to achieve these values, 
and the existence of dark matter and its time variation
cannot explain $d{\rm AU}/dt$.

As mentioned in the previous section, some attempts were made to 
show the secular increase in AU. However, the origin of 
$d{\rm AU}/dt$ is presently far from clear. As the one possibility, 
it is believed that the most plausible reason for the origin of
$d{\rm AU}/dt$ is the lack of calibrations of 
internal delays of radio signals within spacecrafts. Nevertheless, 
as other unexplained anomalies discovered in the solar system, 
$d{\rm AU}/dt$ may be attributed to the fundamental property of
gravity, therefore, this issue should be explored in terms of all
possibilities.

Though it is currently impossible to detect the evidence of dark 
matter from light propagation, some planned space missions, 
especially ASTROD, are aimed to achieve a clock stability of
$10^{-17}$ over a travel time of $1000 ~[{\rm s}]$ \citep{ni2007}. 
Improvement in both the laser ranging technique and the clock stability 
may enable us to observe the trace of dark matter, if it really 
exists in the solar system. For this purpose, it is very important 
subjects to develop a rigorous light propagation model. 

In particular, since it is not easy to analytically
calculate the time-dependent null geodesic equation, in this study, 
we integrated (\ref{metric6}) assuming the simple linear combinations 
of each effect. However, from the theoretical point of view and some 
astronomical and astrophysical applications such as formulation of 
the cosmological gravitational lensing in the expanding 
background, it is noteworthy to develop a method to 
analytically compute the time-dependent geodesic equation.
\section*{Acknowledgments}
We would like to appreciate the anonymous referee for 
fruitful comments. We also acknowledge Prof. G. A. Krasinsky 
for providing information about and comments on the AU issue. 
This work was partially supported by the Ministry of Education, 
Science, Sports and Culture, Grant-in-Aid, No. 21740193.

\end{document}